\documentclass[12pt]{article}
\usepackage{color}
\usepackage{amsbsy}
\usepackage{amsmath}
\usepackage{amssymb}
\usepackage{latexsym}
\usepackage{comment}
\usepackage{graphicx}
\usepackage{amsfonts}
\usepackage{url}
\usepackage{hyperref}
\usepackage[hyphenbreaks]{breakurl}
\usepackage{enumitem}
\usepackage{bbm}

\usepackage[authoryear]{natbib}

\usepackage{geometry}

\def\singlespace{\def\baselinestretch{1}\@normalsize}

\renewcommand{\baselinestretch} {1.2}
\makeatletter 
\def\singlespace{\def\baselinestretch{1}\@normalsize}

\@addtoreset{equation}{section}

\renewcommand{\thefootnote}{\fnsymbol{footnote}}

\newtheorem{theorem}{Theorem}

\newcommand{\norm}{\vert\kern-0.3ex\vert\kern-0.3ex\vert} % norm of random variable (with three vertical lines)
\newcommand{\Bignorm}{\Big\vert\kern-0.4ex\Big\vert\kern-0.4ex\Big\vert} % Big norm of random variable (with three vertical lines)

\def\tilde{\widetilde}
\def\hat{\widehat}
\def\nullH0{\mathrm{null}}

\def\vv{1.28}

\begin{document}

\title{Adjusting the Benjamini-Hochberg method for controlling the false discovery rate in knockoff assisted variable selection}
%\runtitle{Specification  Tests for Covariance Structures}
%\thankstext{T1}{Footnote to the title with the ``thankstext'' command.}
\author{Sanat K. Sarkar\footnotemark[1]
\ and    Cheng Yong Tang \footnotemark[1]  }
\renewcommand{\thefootnote}{\fnsymbol{footnote}}

\footnotetext[1]{Department of Statistical Science,
Temple University,
1810 Liacouras Walk,
Philadelphia, Pennsylvania 19122-6083,
U.S.A. Emails: sanat@temple.edu, yongtang@temple.edu}

\date{}
\maketitle
\renewcommand{\baselinestretch} {\vv}
\normalsize

\begin{abstract}
The knockoff-based multiple testing setup of \cite{Barber2015} for variable selection in multiple regression where sample size is as large as the number of explanatory variables is considered. The method of \cite{Benjamini1995} based on ordinary least squares estimates of the regression coefficients is adjusted to the setup, transforming it to a valid $p$-value based false discovery rate controlling method not relying on any specific correlation structure of the explanatory variables. Simulations and real data applications show that our proposed method that is agnostic to $\pi_0$, the proportion of unimportant explanatory variables, and a data-adaptive version of it that uses an estimate of $\pi_0$ are powerful competitors of the false discovery rate controlling method in \cite{Barber2015}.
\end{abstract}

\section{Introduction} Variable selection, which is an integral part of statistical model building, has become increasingly important in analyzing data from modern scientific investigations. See \cite{Desboulets2018} for a review of numerous state-of-the-art variable selection procedures over a wide range of model structures. Discovery of important variables under multiple linear regression setting guided by a method controlling the false discovery rate (FDR) \citep{Benjamini1995} can lead to a more interpretable and parsimonious model and help scientists to better understand how certain factors influence an outcome.

Let us consider the classical linear regression model:
\begin {eqnarray}\label{Model1} Y = X \beta + \epsilon,\label{eq:lm}\end {eqnarray} where $Y \in \mathbb{R}^{d}$ is the response vector, $X = (X_1, \ldots, X_d)\in \mathbb{R}^{n \times d}$ is the design matrix of rank $d \le n$ with its columns representing known observation vectors on $d$ explanatory variables, $\beta = (\beta_1, \ldots, \beta_d)\in \mathbb{R}^{d}$ is the unknown vector of regression coefficients, and $\epsilon \sim \mathcal{N}_d(0, \tau^2 I_d)$, with unknown $\tau^2$, is the Gaussian noise.

Variable selection under this model can be framed as a multiple testing problem with $H_j: \beta_j = 0$ being the null hypothesis tested against $\beta_j \neq 0$, simultaneously for $j=1, \ldots, d$. The explanatory variables corresponding to the rejected null hypotheses according to a multiple testing method based on appropriately chosen marginal test statistics obtained from some estimates of the regression coefficients can be selected/discovered as the important variables. The most appropriate choice of these test statistics are the ones that are based on the ordinary least squares (OLS) estimate of $\beta$ under Model \ref{Model1}, which is $\hat{{\beta}}_{\textrm{OLS}} = \Sigma^{-1} {X}^{\prime}{Y} \sim \mathcal{N}_d(\beta, \tau^2 \Sigma^{-1})$, where $\Sigma= {X}^{\prime}{X}$, since each of these marginal tests will be optimal \citep{LR2005}.

The FDR, which is defined as follows: \begin {eqnarray} \textrm{FDR} = \mathbb{E}\left ( \frac{V}{\max(R,1)}\right ), \nonumber \end {eqnarray} with $V$ being the number of falsely selected/discovered variables and $R$ being the total number of selected variables, is a powerful measure of potential type I errors associated with such a selection procedure. The method of \cite{Benjamini1995}, popularly known as the BH method, is one of the most commonly used methods for controlling this error rate.
%under similar multiple testing scenarios. Unfortunately, however, the BH method based on $\hat{{\beta}}_{\textrm{OLS}}$ does not work in general for variable selection in the current non-asymptotic setting. It requires certain dependency of the marginal test statistics. More specifically, the BH method
Let $P_j$ be the $p$-value associated with $H_j$ and $P_{(j)}$ be the $j$th value when the $p$-values are sorted from smallest to largest. Then, the level $\alpha$ BH method  rejects $H_j$ for each $j$ such that $P_j \le P_{(R)}$, where $R= \max\{j: P_{(j)} \le j \alpha/d\}$, provided the maximum exists; otherwise,  it rejects none. It provably controls the FDR under positive dependency of the underlying test statistics that is characterized by the condition of positive regression dependence on subset of null test statistics or by its stronger version of multivariate totally positive of order two. \cite{Benjamini2001} introduced the positive regression dependence condition before providing a proof of the FDR control of the BH method under it; see also \cite{Blanchard2008}, \cite{Finner2007} and \cite{Sarkar2008b} for alternative proofs. \cite{Karlin1980} introduced the multivariate totally positive dependence condition, and under this condition, \cite{Sarkar2002} strengthened the work of \cite{Benjamini2001} by giving alternative proof of this FDR control for a more general class of multiple testing methods.

For a multivariate Gaussian distribution with mean vector $\mu$ and positive definite covariance matrix $\Gamma$, the positive regression dependency holds if $\Gamma_{ij} \ge 0$, for all $i \neq j$, \citep{Benjamini2001} and the multivariate total positivity holds if $(\Gamma^{-1})_{ij} \le 0$, for all $i \neq j$, \citep{Karlin1980}. These conditions  trivially hold for $\Gamma$ when it is diagonal. For a non-diagonal $\Gamma$, these conditions can often be checked or justified in the context of multiple testing when the alternatives are all either right- or left-sided, but they are not generally satisfied for two-sided alternatives like what we have in variable selection. These conditions also hold for the absolute valued multivariate $t$ when the associated correlation matrix is diagonal.

So, the application of the BH method based on $\hat{\beta}_{\textrm{OLS}}$ to control the FDR in variable selection, unfortunately, is questionable, except in the very unusual case when the $X_j$'s are orthogonal.
%The Benjamini-Hochberg (BH) method based on $p$-values calculated from the ordinary least squares (OLS) estimates of the regression coefficients would be most appropriate and a powerful tool for the underlying testing problem. Unfortunately, it does not work in the current setting, since its FDR control requires certain type of positive dependence condition among the test statistics (\cite{Benjamini2001} and \cite{Sarkar2002}) that is not generally exhibited by the correlation matrix given by the Gram matrix, unless it is diagonal.
\cite{Barber2015} introduced an ingenious approach to controlling the FDR in variable selction.
%in multiple linear regression when sample size is as large as the number of explanatory variables, in which the final selection of important variables is indeed guided by methods that control the FDR.
The novelty of their approach lies in the construction of knockoff copy of the design matrix allowing formation of a framework for the underlying multiple testing problem that produces distribution-free and powerful FDR controlling algorithms. These are flexible and widely applicable, since $p$-value calculations and knowledge of any specific correlation structure of the explanatory variables are no longer required.

Specifically, \cite{Barber2015} first considered the case when $n \ge 2d$, and constructed $\tilde{X}= (\tilde{X}_1, \ldots, \tilde{X}_d)$, the knockoff copy of $X= (X_1, \ldots, X_d)$, satisfying
\begin{eqnarray} \label{Knockoff1} \tilde{{X}}^{\prime} \tilde{{X}} = {\Sigma} \quad \mbox{and} \quad  \tilde{{X}}^{\prime} {X} = {\Sigma} - {D}, \end{eqnarray}
where ${D} = \mathrm{diag}\left\{{s} \right \}$, with ${s} \in \mathbb{R}_{+}^d$ being such that $2{\Sigma}- {D}$ is positive definite. The following is an explicit representation of knockoff of ${X}$:
\begin {eqnarray} \label{knockoff2} \tilde{{X}} = {X} {\Sigma}^{-1}({\Sigma} - {D} ) + \tilde{{U}}(2{D} - {D} {\Sigma}^{-1} {D})^{\frac{1}{2}}, \end {eqnarray} where the columns of $\tilde{{U}} \in \mathbb{R}^{n \times d}$ are orthonormal and orthogonal to the column space of ${X}$ and $(2{D} - {D} {\Sigma}^{-1} {D})^{\frac{1}{2}}$ is positive square root of $2{D} - {D} {\Sigma}^{-1} {D}$, which exists due to the above condition on ${D}$.

Once the knockoff $\tilde{{X}}$ is constructed, the augmented version of the model in \eqref{Model1}, with ${X}$ replaced by (${X}, \tilde{{X}}$),
%with ${X}$ replaced by $({{X}}, \tilde{{X}})$ and ${\beta}$ by $({\beta}, \tilde{{\beta}})^{\prime}$,
is used to estimate the pair of regression coefficients associated with the $j$th explanatory variable and its knockoff copy, for each $j=1, \ldots, d$, using an estimation method. Given a statistic that tends to have larger positive values when $\beta_j \neq 0$ than when $\beta_j=0$, and one would have used under the model in (1) to test $H_j$ against $\beta_j \neq 0$, i.e., the $j$th explanatory variable is important or not, there will be $d$ pairs of such statistics $(Z_j,\tilde{Z}_j)$, $j=1, \ldots d$, when the augmented model is fitted. Among other choices, the test statistic $Z_j$ can be defined as the value of the penalty parameter associated with the Lasso solution path at which the $j$th variable
first enters the Lasso model.
%or the differences between the absolute values of the estimates corresponding $X_j$ and $\tilde X_j$. %is such a statistic among others.
\cite{Barber2015} propose their FDR controlling methods using $W_j= (Z_j \vee \tilde{Z}_j)[2 \mathbbm{1}(Z_j > \tilde{Z}_j) - 1]$.
%for multiple testing of $H_j$ against $\beta_j \neq 0$, for $j=1, \ldots, d$, since a large positive value of $W_j$ would indicate that variable $X_j$ is genuinely important and belongs in the model. Although, as noted by them, these $W_j$'s can be constructed in other possible ways as long as certain sufficiency and antisymmetry conditions hold. These conditions ensure a crucial property of the $W_j$'s, which is, that they are iid random variables for the null hypotheses, and they are independent of the $|W_j|$ for all $j$ and of $\textrm{sign}(W_j)$ for all $j$ that correspond to the non-null hypotheses. \cite{Barber2015} considered two different notions of FDR, one is as defined in (\ref{Eqn1}) and the other is a slight variation of it,
More specifically, they have considered rejecting $H_j$ if $W_j > T$, with a data-dependent threshold $T$ being determined as follows:
\begin {eqnarray} \label{eq:bct}
T = \min \left \{ t \in \mathcal{W}: \frac{1+ \# \{j: W_j \le -t \}}{\# \{j: W_j \ge t\} \vee 1} \le \alpha \right \} \end{eqnarray} or $T = +\infty$ if this set is empty, where $\mathcal{W} = \{|W_j|: j =1, \ldots, d\} \setminus \{0\}$, to control the FDR at a pre-specified level $\alpha$ without any distributional assumptions. They have considered using a slightly different data-dependent threshold $T^*$ as well, which is given by
\begin {eqnarray*} T^* = \min \left \{t \in \mathcal{W}: \frac{\# \{ j: W_j \le -t\}}{ \# \{ j: W_j \ge t \} \vee 1} \le \alpha \right \} \end{eqnarray*} or $T^* = +\infty$ if this set is empty. This, however, controls a slightly different measure called ``modified FDR" at $\alpha$. Intuitive extensions of these methods when $d < n \le 2d$ have also been given in \cite{Barber2015}.

Since  \cite{Barber2015}, a considerable amount of research has taken place in the knockoff domain, with papers written on variable selection centered around the knockoff-based FDR controlling algorithms in \cite{Barber2015}, notable among which are \cite{Barber2019}, \cite{Candes2018}, \cite{Barber2020}, \cite{Romano2019}, \cite{Sesia2018}, and \cite{Janson2021}. The BH method is often considered as a competing FDR controlling method among those that are knockoff-based  \citep{Candes2018,Barber2019,Xing2019}, even though its validity as an FDR controlling method without being adjusted to the underlying correlation structure is uknown. This drives our motivation to the current research on adjusting the BH method to the underlying knockoff setting to make it a valid, powerful FDR controlling method that does not rely on any specific correlation structure of the explanatory variables.

%In this article, we take a different approach. It aims at making use of the knockoff copy of $X$ in developing $p$-value based BH type methods that fully capture $\Sigma$ and provably controls the FDR without relying on any specific correlation structure of the explanatory variables. The proposed methods are presented in the next section.

\section{Proposed methods}

\subsection {Case I: $n \ge 2d$} The development of the proposed methods in this case is driven by our observation %from (\ref{Eqn:Knockoff1}) and (\ref{Knockoff2})
 that the knockoff $\tilde{{X}}$, once created, yields the following two distinct estimators of ${{\beta}}$:
\begin{eqnarray*}
\hat{{\beta}}_1 & = & (2{\Sigma} - {D} )^{-1} ({X} + \tilde{{X}})^{\prime}{Y} =   \hat{{\beta}}_{\textrm{OLS}} + (2{\Sigma} - {D} )^{-1}(2{D} - {D} {\Sigma}^{-1} {D})^{\frac{1}{2}}\tilde{{U}}^{\prime}{Y} \end {eqnarray*}
and
\begin{eqnarray} \label{Eqn:Betas}\hat{{\beta}}_2 & = & {D}^{-1} ({X} - \tilde{{X}})^{\prime}{Y} = \hat{{\beta}}_{\textrm{OLS}} - {D}^{-1}(2{D} - {D} {\Sigma}^{-1} {D})^{\frac{1}{2}}\tilde{{U}}^{\prime}{Y}, \end{eqnarray}
%with $\hat{{\beta}}_{\textrm{OLS}}=\Sigma^{-1} {X}^{\prime}{Y}$ being the ordinary least squares estimator of ${\beta}$ under the model in \eqref{Model1},
which are independently distributed under Gaussian noise assumption as $\mathcal{N}_{d}( {\beta}, 2\tau^2(2{\Sigma} - {D})^{-1} )$ and  $\mathcal{N}_{d} ({\beta}, 2\tau^2{{D}}^{-1})$, respectively.
%These two estimator, of course, reduce to $\hat{{\beta}}$ when ${\Sigma}={D}$.
Thus, the knockoff allows creating two independent settings for multiple testing of the same $H_j$'s, in one of which the marginal test statistics are uncorrelated. Thus, we have a novel framework for the underlying testing problem where the BH method can work as a valid FDR controlling method, either independently by itself or integratively with the Bonferroni.

More specifically, let us define the vectors of test statistics
\begin{eqnarray}
{T}_1 = (T_{11}, \ldots, T_{1d})^{\prime} = \frac{1}{\hat{\tau}\sqrt{2}}[\textrm{diag}\left\{(2{\Sigma}-{D})^{-1} \right \}]^{-\frac{1}{2}} \hat{{\beta}}_1 \nonumber
\end{eqnarray}
and
\begin{eqnarray}
{T}_2 = (T_{21}, \ldots, T_{2d})^{\prime} = \frac{1}{\hat{\tau}\sqrt{2}}{D}^{\frac{1}{2}} \hat{{\beta}}_2. \nonumber \end{eqnarray}
The estimated noise variance $\hat{\tau}^2$, obtained from the linear regression model with the augmented design matrix $({X},\tilde{{X}})$, is distributed independently of $\hat{{\beta}}_1$ and $\hat{{\beta}}_2$ as $\tau^2 \chi_{n-2d}^2/(n-2d)$. From these, we obtain the following pair of $p$-values
\begin{eqnarray} \label{eq:t}
P_j^{(t_1)} = \mathbb{P}(t_{n-2d}^2 > T_{1j}^2) \quad \mbox{and} \quad P_j^{(t_2)} = \mathbb{P}(t_{n-2d}^2 > T_{2j}^2), \nonumber
\end{eqnarray}
with $t_{n-2d}$ being the $t$ distribution with $n-2d$ degrees of freedom, each of which provides a test for $H_j$, for $j=1, \ldots, d$.  Since $\hat\beta_2\sim \mathcal{N}_{d} ({\beta}, 2\tau^2{{D}}^{-1})$ with a diagonal $D$, the BH method based on ${P}^{(t_2)}= (P_1^{(t_2)}, \ldots, P_d^{(t_2)})$ itself can now be used \citep{Benjamini2001, Sarkar2002}, or it can be combined with the Bonferroni method based on ${P}^{(t_1)}= (P_1^{(t_1)}, \ldots, P_d^{(t_1)})$, to control the FDR.
%The BH method can now be used, either by itself based on ${P}^{(t_2)}= (P_1^{(t_2)}, \ldots, P_d^{(t_2)})$ [see, for instance, \cite{Benjamini2001} and  \cite{Sarkar2002}] or by combining it with the Bonferroni method based on ${P}^{(t_1)}= (P_1^{(t_1)}, \ldots, P_d^{(t_1)})$, to control the FDR.
In other words, one can consider screening the potentially significant hypotheses by applying Bonferroni-type common thresholding to  ${P}^{(t_1)}$ and then incorporating the corresponding $p$-values into the BH method based on ${P}^{(t_2)}$ to determine which of these should be finally rejected to achieve the desired control over the FDR. Morover, the BH method can be used in its original form that is agnostic to $\pi_0$, the proportion of true nulls, or in its data-adaptive form that utilizes an estimate of $\pi_0$ based on the existing data. Thus, we have two proposed FDR controlling methods, as stated below. %These and the theorems stating their FDR controlling properties are presented below with proofs of the theorems being provided in Supplementary Material.
%The rest of the paper is organized as follows. The proposed methodological framework and  the underlying theory are elaborated in Section \ref{s3}.  The results of an extensive simulation study  examining the performances of our proposed procedures are reported in Section \ref{s4}, followed by a real data example in Section \ref{s5}. We conclude this article with some additional remarks and discussions in Section \ref{s6}.

\vskip 6pt
{\it Method 1.} Bonferroni-BH.
\begin{itemize}
\item [Step 1] Given $\alpha \in (0,1)$, the level at which the FDR is to be controlled, let
\begin {eqnarray}\widetilde{P}_{j} = \left \{
\begin{array}{lll}
1 & \mbox {if} & P_j^{(t_1)} > \sqrt{\alpha}\\
P_j^{(t_2)} & \mbox {if} &  P_j^{(t_1)} \le \sqrt{\alpha}
\end{array},
\right. \nonumber \end {eqnarray}  for $j=1, \ldots, d$.
%\underline{} \
%$\widetilde{P}_{j} = 1- \mathbbm{1}(P_j^{(t_1)} \le \sqrt{\alpha})(1-P_j^{(t_2)})$, for $j=1, \ldots, d$. \
%\item [Step 2.] Let $C^{(-1)}(\alpha | \lambda)$ be such that $C(C^{(-1)}(\alpha | \lambda)|\lambda)Determine $\alpha_1(\lambda) \le \cdots \le \alpha_d(\lambda)$ such that $$ C(\alpha_j(\lambda), \lambda)= j \alpha/d, \; \mbox{for} \; j=1, \ldots. d. $$ \
\item [Step 2] With $\widetilde{P}_{(1)} \le \cdots \le \widetilde{P}_{(d)}$ as the ordered versions of the $\widetilde{P}_{j}$'s, find $$\widetilde{R} = \max \left \{i: \widetilde{P}_{(i)} \le i \sqrt{\alpha}/d \right \},$$ provided the maximum exists; otherwise, let it be $0$.\
\item [Step 3] Reject $H_{(j)}$, the hypothesis corresponding to $\widetilde{P}_{(j)}$, for all $j \le \widetilde{R}$.
\end{itemize}

\begin{theorem}\label{Theorem1} Method 1 controls the FDR at $\pi_0\alpha$ under \eqref{eq:lm}. \end{theorem}

 %As in the regular BH method,
The Bonferroni-BH method can be adapted to the existing data through estimating $\pi_0$, thereby tightening its FDR control at $\alpha$ and hence potentially improving its power. There are several different  choices for this estimate; see, for example, \cite{Blanchard2009} and \cite{Sarkar2008b}. Among these, the one in \cite{Storey2004} is frequently chosen. We will consider using this estimate to introduce the following adaptive version of the Bonferroni-BH.

\vskip 6pt
{\it Method 2.}
Adaptive Bonferroni-BH.
\begin{itemize}
\item [Step 1.] Consider the following estimate of $\pi_0$ based on ${P}^{(t_2)}$:
\begin {eqnarray} \hat{\pi}_{0}^{(t_2)} = \frac{d - \sum_{j=1}^d \mathbbm{1}(P_j^{(t_2)} \le \eta) +1}{d(1-\eta)},  \label{eq:adp} %\nonumber
\end {eqnarray}
for some fixed $\eta \in (0,1)$.\
\item [Step 2.] For $j=1, \ldots, d$,  let \begin {eqnarray} {P}_{j}^{*} = \left \{
\begin{array}{lll}
1 & \mbox {if} & P_j^{(t_1)} > \sqrt{\alpha}\\
\hat{\pi}_{0}^{(t_2)}P_j^{(t_2)} & \mbox {if} &  P_j^{(t_1)} \le \sqrt{\alpha}.  \\
\end{array}
\right. \nonumber \end {eqnarray}

\item [Step 3] With ${P}_{(1)}^{*} \le \cdots \le {P}_{(d)}^{*}$ as the ordered versions of the ${P}_j^*$'s, find ${R}^{*} = \max \{i: {P}_{(i)}^{*} \le i \sqrt{\alpha}/d \},$ provided the maximum exists; otherwise, let it be $0$.\
\item [Step 4] Reject $H_{(j)}$, the hypothesis corresponding to ${P}_{(j)}^{*}$, for all $j \le {R}^{*}$.
\end{itemize}

\begin{theorem}
\label{Theorem2} Method 2 controls the FDR at $\alpha$ under \eqref{eq:lm} when $n$ is sufficiently large, so that $\hat {\tau}^2 \approx \tau^2$.
\end{theorem}

{\it Remark} \rm 1. (i) When $\boldsymbol{\Sigma}$ is diagonal, Methods 1 and 2 reduce, respectively, to the original BH and a data-adaptive version of it based on the ordinary least squares estimate of $\boldsymbol{\beta}$. So, our proposed methods can be regarded as  knockoff-adjusted versions of the BH method in its original and data-adaptive forms.

(ii) When $\tau^2$ is known or $\hat\tau^2 \approx \tau^2$, with $n$ being sufficiently large, the $p$-values in our proposed Methods 1 and 2 are determined from test statistics that are distributed (or approximately distributed) as squares of normals with unit variance, and $P^{(t_1)}$ and $P^{(t_2)}$ are independent.

(iii) Methods 1 and 2 are clearly not meant for variable selection in high-dimensional settings ($d \gg n$), even though they act like screening-and-testing type algorithms commonly employed in such settings with the data being split into two parts \citep{Barber2019, Meinshausen2009, Wasserman2009}.
Of course, the algorithms here for low-dimensional settings are more tractable, since the $p$-values in the testing step are independent, or conditionally independent given $\hat {\tau}^2$, of those in the screening step, unlike in the high-dimensional case.

(iv) We could have defined our knockoff-assisted variable selection without going through the screening step, since in the testing step we can have the tightest possible control of the FDR (in some instances, as noted below). However, supplementing it with a screening step, as we did while defining Methods 1 and 2, can enhance their power. Highly important variables with relatively small $p$-values are more probable to be selected as they are more likely to be picked up among others sharing some form of local dependency in the screening step for possible final selection. On the adaptability of our proposed methods to high-dimensional settings, we have made some remarks about it in Section 6.

(v) It is important to note that Methods 1 and 2 are special cases of a scenario %of those that arise in the context of a more general setting, %not necessarily restricted to that in variable selection,
that  we refer to as paired-multiple testing setting. In this general setting, we have a pair of $p$-values, $(P_{j}^{(x)}, P^{(y)}_{j})$, each of which provides a test for a hypothesis $H_j$ associated with the same parameter shared by a pair of generic random variables ($X_j, Y_j$), for $j=1, \ldots, d$,  and the problem is to simultaneously test these $d$ null hypotheses using both sets of $p$-values $ {P}^{(x)} = (P_{1}^{(x)}, \ldots, P_{d}^{(x)})$ and ${P}^{(y)} = (P_{1}^{(y)}, \ldots, P_{d}^{(y)})$ subject to a control of the FDR at $\alpha$.
The generalizations of Methods 1 and 2 are referred to as Methods S1 and S2, respectively, as elaborated in
Supplementary Material.
%We give our proofs of Theorems 1 and 2 in Supplementary Material, having stated them for Methods 1 and 2, respectively, in
These general forms use a tuning parameter $\lambda$, instead of $\sqrt{\alpha}$, in their screening steps, which shows the flexibility that Methods 1 and 2 enjoy in their applications.

(vi) Theorems 1 and 2  are  thus special cases of Theorems S1 and S2 (corresponding to the aforementioned Methods S1 and S2, respectively), whose statements and proofs are given in Supplementary Material.
%Their FDR controls are established, respectively, at $\pi_0 \alpha$ and $\alpha$
Our theory is established
under some general assumptions and conditions on (${P}^{(x)}, {P}^{(y)}$), broadening their scopes beyond the linear models. %of Methods 1 and 2.
Following our proofs of these general theorems, we will explain why the assumptions and conditions associated with these general methods cover model   \eqref{eq:lm}  in the context of knockoff-assisted variable selection.
More specifically, Method 1 provably controls the FDR  at  $\pi_0 \alpha$ even when the components of ${P}^{(y)}$, instead of being just independent, as in the current knockoff-assisted variable selection setting, are positively regression dependent on the null $p$-values (formally defined in Supplementary Material), and Method 2 controls the FDR at  $\alpha$ for a certain class of estimates containing the one originally considered in defining it.  As noted from these proofs, if $\tau^2$ is known or $\hat\tau^2 \approx \tau^2$ with $n$ being sufficiently large, the FDR of Method 1 exactly equals $\pi_0 \alpha$, and hence it provides the tightest possible control of the FDR in the sense of its FDR being exactly equal to the level when there are no signals.

\subsection{Case II: $d < n \le 2d$} We can proceed as in \cite{Barber2015} to propose our methods in this case. Specifically, we can augment the response vector $Y$ with a $(2d-n)$-dimensional observation vector $Y^*$ randomly taken from $\mathcal{N}_{2d-n}(0,\tau^2I_{2d-n})$ if $\tau^2$ is known or from $\mathcal{N}_{2d-n}(0,\hat{\tau}_\text{OLS}^2I)$, where $\hat{\tau}_\text{OLS}^2 = \| Y-X\hat{\beta}_\text{OLS}\|^2/(n-d)$, if $\tau^2$ is unknown but $n-d$ is large so that $\hat{\tau}_\text{OLS}^2 \approx \tau^2$. Augmenting $X$ row-wise with $2d - n$ rows of zeros, we then have the following multiple linear regression model
\begin{eqnarray} \left(
    \begin{array}{c}
      Y\\
      Y^* \\
    \end{array}
  \right) = \left ( \begin{array}{c}
      X\\
      0 \\
    \end{array}
  \right ) \beta + \epsilon, \nonumber \end{eqnarray} with $(Y^{\prime}, {Y^*}^{\prime})^{\prime} \in \mathbb{R}^{2d}$ as the response vector, $(X^{\prime},0^{\prime})^{\prime} \in \mathbb{R}^{2d \times d}$ as the design matrix of rank $d$ with its columns representing known observation vectors on $d$ explanatory variables, $\beta$ as the unknown vector of regression coefficients, and $\epsilon \sim \mathcal{N}_{2d}(0, \tau^2 I)$, exactly or approximately depending on whether $\tau^2$ is known or $n-d$ is large letting $\hat{\tau}_\text{OLS}^2 \approx \tau^2$.
 We can apply our proposed knockoff-assisted variable selection under this model %using Method 1 or 2 with $n=2d$ and $p$-values
as mentioned in Remark 1(ii).

 \section{Simulations} \label{s4}

This section presents the results of an extensive simulation study we conducted to numerically investigate the performances of our proposed Bonferroni-BH method and its data-adaptive version in terms  of FDR control and power (expected proportion of correct discoveries). The data were generated from \eqref{eq:lm}.  The components of the model matrix were generated from correlated standard normal random variables having an auto-regressive structure of order 1 with coefficient 0.5.  The model error was generated independently from standard normal distribution.
%With the knockoff method of \cite{Barber2015} being the relevant competitor of our proposed methods, we benchmarked the performances of all these methods against the original BH method based on the $\hat{{\beta}}_2$ statistic, which is known to control the FDR exactly at $\pi_0 \alpha$.
We attempted the following different settings in our simulation study:
\begin{itemize}
\item Sample size: $n\in\{100, 200, 500,1000\}$;
\item  The number of explanatory variables: $d\in\{0.1n,0.2n,0.3n,0.4n\}$;
\item The number of important variables: $k\in\{0,\lceil 0.1d\rceil, \lceil 0.2d \rceil, \lceil 0.3d \rceil, \lceil 0.4d \rceil, \lceil 0.5d \rceil\}$  where $\lceil \cdot\rceil$ is the ceiling function;
\item Signal strength: $\beta_1=\dots=\beta_k=a\in \{2,4,6,8,10\}$ for different amplitude $a$;
\item FDR level: $\alpha\in \{0.05,0.1\}$;
%\item The thresholding parameter for Bonferroni: $\lambda=\sqrt{\alpha}$;
\item The tuning parameter for estimation of true nulls: $\eta=0.5$ for \eqref{eq:adp}.
\end{itemize}
For each setting, the simulations were repeated 500 times. The knockoff copy was created following the method of \cite{Barber2015} by implementing the ``Equi-Correlated Knockoffs'' available therein by choosing ${D} = \mathrm{diag}\left\{{s} \right \} = \mathrm{diag} \left\{s,\dots, s \right \}$ with $s=2\xi_{\text{min}}({\Sigma})\lor 1$, where $\xi_{\text{min}}({\Sigma})$ is the smallest eigenvalue of ${\Sigma}$.

%When $k=0$, all null hypotheses are true so that any discovery is false.
We compared the two proposed methods with the knockoff method of \cite{Barber2015} (that uses the data-dependent threshold $T$ in \eqref{eq:bct} with the $W_j$ being calculated from the Lasso penalty parameter). Also included in this comparison is the original BH method based on just $\hat{\beta}_2$ that  also controls the FDR.

\begin{figure}
\begin{tabular}{ccc}
\includegraphics[width=0.3\textwidth]{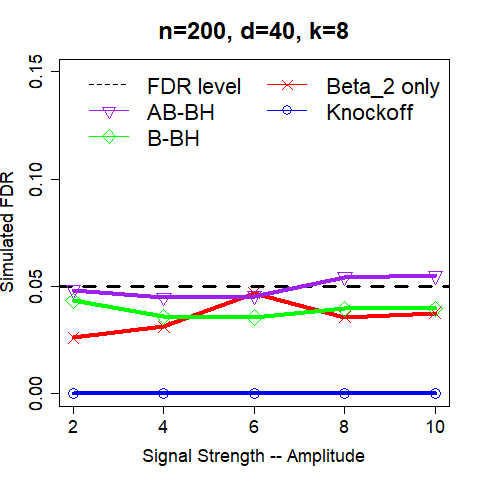}&
\includegraphics[width=0.3\textwidth]{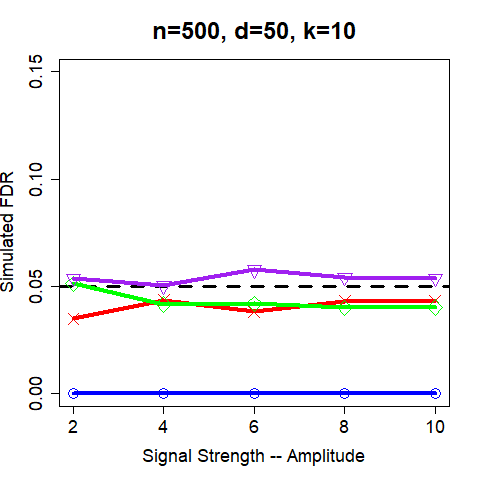}&
\includegraphics[width=0.3\textwidth]{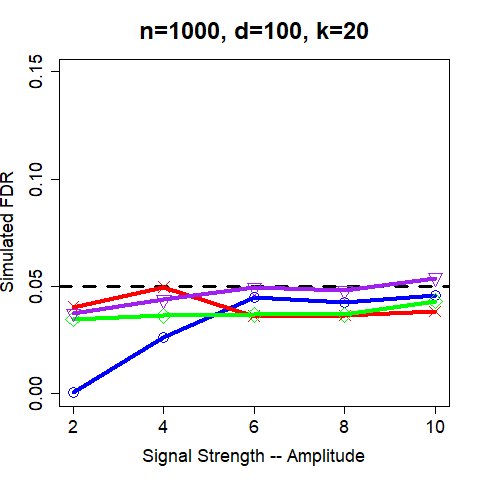}\\
\includegraphics[width=0.3\textwidth]{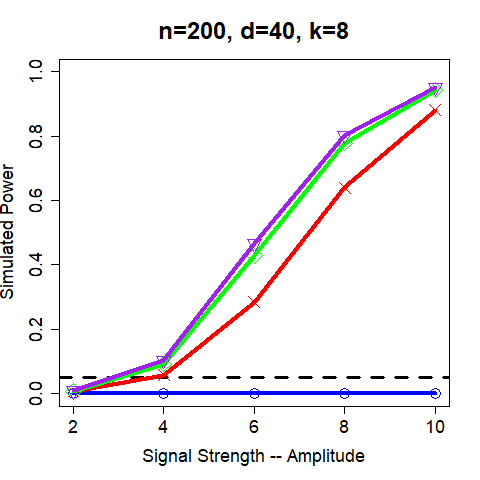}&
\includegraphics[width=0.3\textwidth]{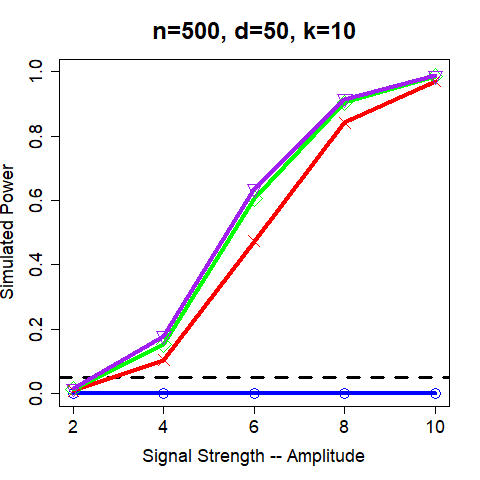}&
\includegraphics[width=0.3\textwidth]{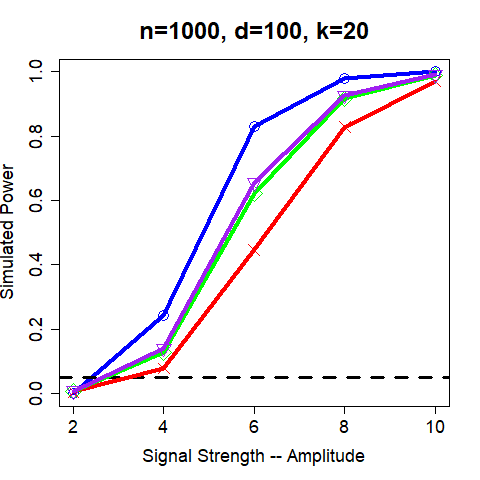}\\
\end{tabular}
\caption{Simulated FDR (1st row) and power (2nd row) when some null hypotheses are not true; the FDR level is $\alpha=0.05$.
\\
B-BH -- the proposed Bonferroni-BH method; % with $\lambda=\sqrt{\alpha}$;
AB-BH -- the proposed adaptive Bonferroni-BH method; Beta\_2 only refers to the approach using  $\hat{\beta}_2$ only;  Knockoff -- the knockoff based method of \cite{Barber2015}.\\
\label{fig2} }
\end{figure}

\begin{figure}
\begin{tabular}{ccc}
\includegraphics[width=0.3\textwidth]{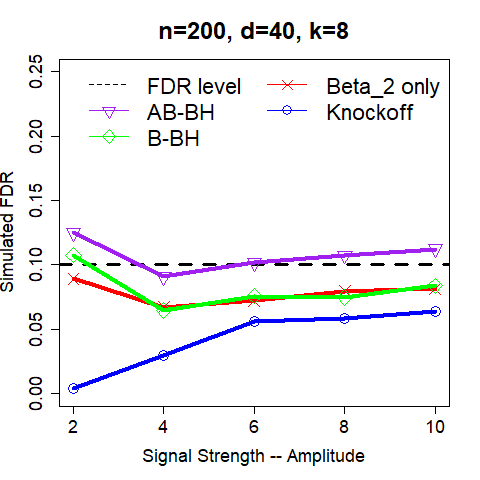}&
\includegraphics[width=0.3\textwidth]{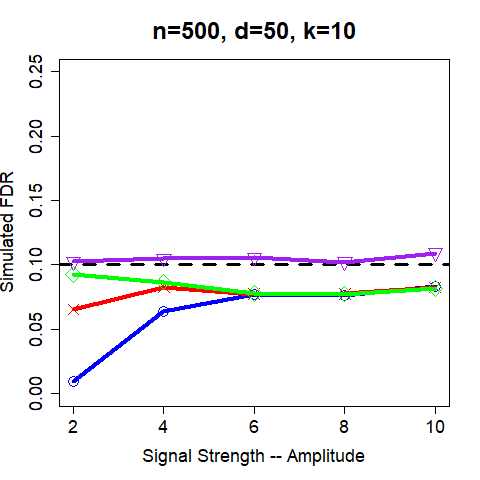}&
\includegraphics[width=0.3\textwidth]{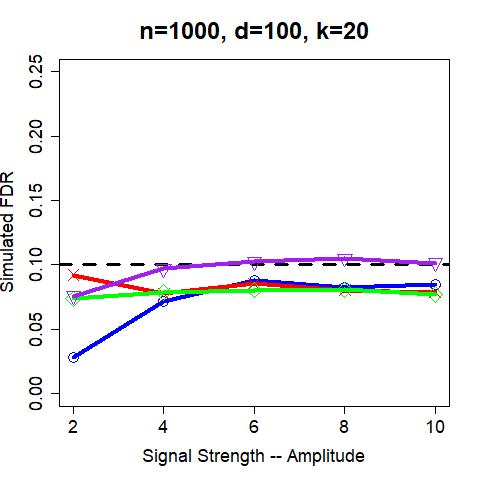}\\
\includegraphics[width=0.3\textwidth]{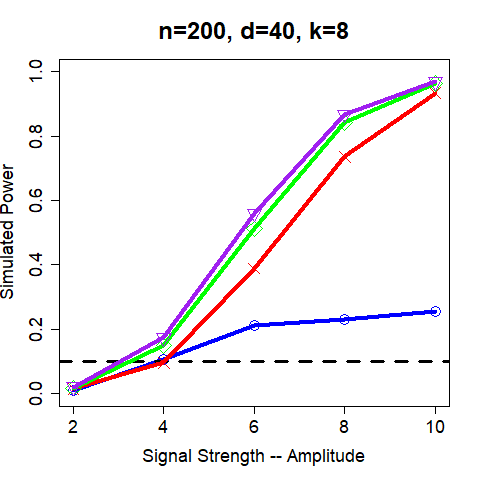}&
\includegraphics[width=0.3\textwidth]{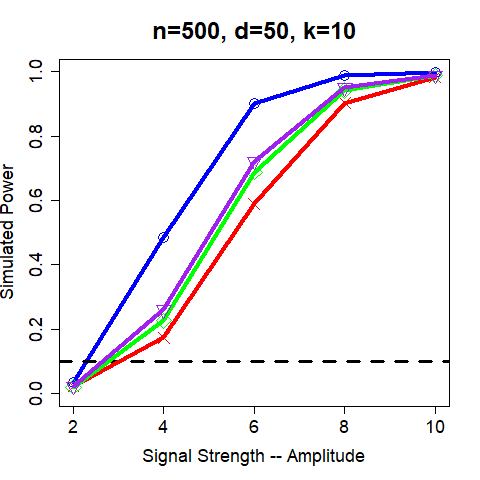}&
\includegraphics[width=0.3\textwidth]{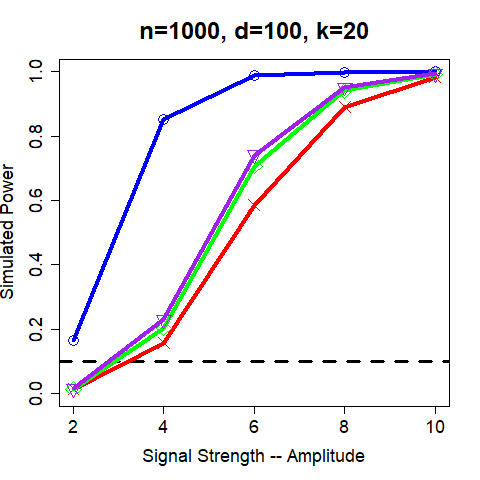}\\
\end{tabular}
\caption{Simulated FDR (1st row) and power (2nd row) when some null hypotheses are not true;  the FDR level is $\alpha=0.1$.
\\
B-BH -- the proposed Bonferroni-BH method; % with $\lambda=\sqrt{\alpha}$;
AB-BH -- the proposed adaptive Bonferroni-BH method; Beta\_2 only refers to the approach using  $\hat{\beta}_2$ only;  Knockoff -- the knockoff based method of \cite{Barber2015}.\\
\label{fig2a} }
\end{figure}

Figures \ref{fig2} and \ref{fig2a} display a set of representative results from our simulation study. They compare the simulated FDRs and powers of all these four methods across different signal strengths, providing a picture of how the signal strength impacts the performances of the methods, respectively, for $\alpha = 0.05$ and $0.10$, when $20$\% of the variables are truly important,  %\sout{extra dimension from the impact of the signal strength is examined} {\color{red}
  %of all the procedures is also examined.
The  first rows in these figures show consistent pattern in the FDR control.
While the FDR is controlled for each of the methods, the knockoff method seems conservative when the signals are relatively weak  and the sample size is not large.  The  second  rows in these figures show very promising power performances of the proposed methods, especially when  the sample size is smaller and the FDR level is $\alpha=0.05$. %lower than $0.10$.
Additionally, the Bonferroni-BH is seen to have better power when adapted to the data via estimating the proportion of true nulls, with both having better power than the $\hat{\beta}_2$-only based BH method. %At a lower FDR level $\alpha = 0.05$, our methods perform very promisingly.

We obtained simulated values of FDR in some other settings for all these four methods across different values of $k$, covering cases from no signal ($k=0$, all nulls are true) to different percentages of signals among the hypotheses. These are displayed in Figures \ref{fig1} and \ref{fig1a} in  Supplementary Material. %\sout{In these two figures, when $k\neq 0$, the signal strength is $a=6$.}
As seen from these figures, while all these methods control the FDR, as expected since it was theoretically established, the Barber-Cand\`es method is seen  overly conservative when there are no or few signals, most often making no discovery, even when the sample size is $1,000$.

%With the results in Figures \ref{fig2} and \ref{fig2a} being for a moderate setting where   $20\%$ of the predictors are contributing,

The overreaching conclusion that can be made from our entire simulation study, including those settings that are not reported here, is that the knockoff method has better performance at larger FDR levels, with denser signals, and for larger sample sizes. Whereas, our proposed Bonferroni-BH and Adaptive Bonferroni-BH perform well at smaller FDR levels, with sparser signals, and for                                                 smaller sample sizes. In other words, our proposed methods can perform quite well relative to the Barber-Cand\`es method in some commonly occurring instances, making them competitive among FDR controlling multiple testing methods for variable selection. %under low-dimensional liner regression.
%\sout{ Moreover, our proposed methods are clearly more appropriate benchmarks than the conventional BH procedure in broad comparisions.}

 %Especially when the signals are sparse with small sample size.

\section{A real data example}\label{s5}

We analyzed the same data set that was considered in \cite{Barber2015}.  This data set has been described and studied in \cite{Rhee2006}, and it is available to the public. The data set contains drug resistance measurements and genotype information from samples of HIV-1.  Separated data sets are available  for resistance, respectively, to protease inhibitors (PIs), to nucleoside reverse transcriptase (RT) inhibitors (NRTIs), and to
nonnucleoside RT inhibitors (NNRTIs). The same data preparing and pre-processing steps as those in \cite{Barber2015} were followed.

We considered analyzing the resistance to protease inhibitors (PIs) data. In this data set, there are 7 drugs. We considered the log-fold increase of lab-tested drug resistance as the response variable, and the same model matrix as that in \cite{Barber2015}. We also compared the selected mutations for various methods with existing treatment-selected mutation (TSM) panels \citep{Rhee2005}. As pointed out in \cite{Barber2015}, the TSM list is a reference consisting of mutations associated with the general PI class of drugs, and so the TSM list is not expected to be specific to the individual drugs.

\begin{figure}
\begin{tabular}{ccc}
\includegraphics[width=0.28\textwidth]{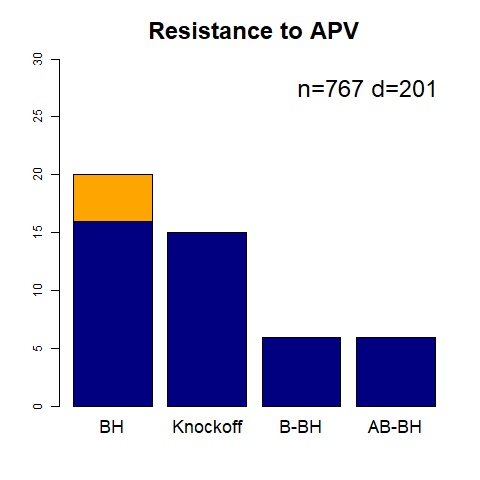}&
\includegraphics[width=0.28\textwidth]{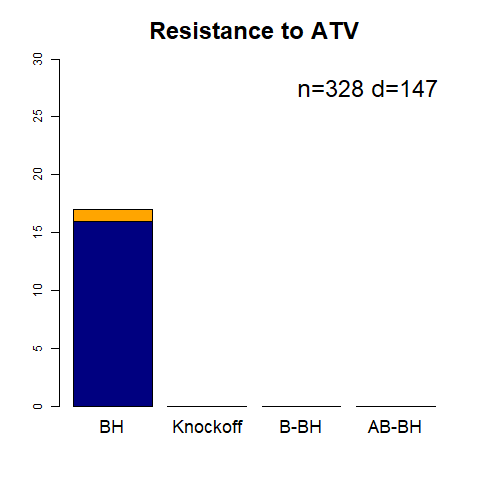}&
\includegraphics[width=0.28\textwidth]{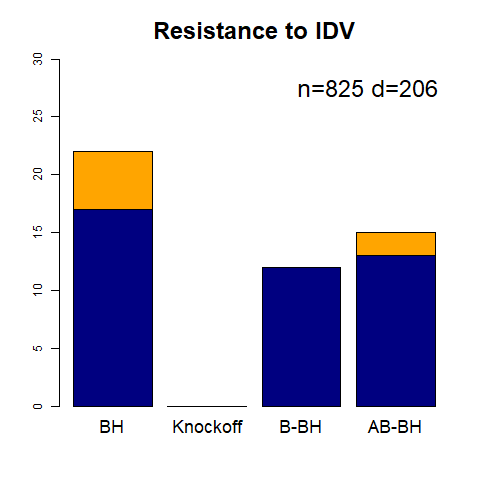}\\
\includegraphics[width=0.28\textwidth]{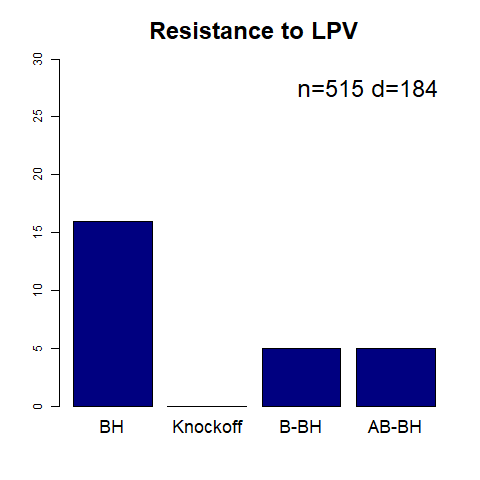}&
\includegraphics[width=0.28\textwidth]{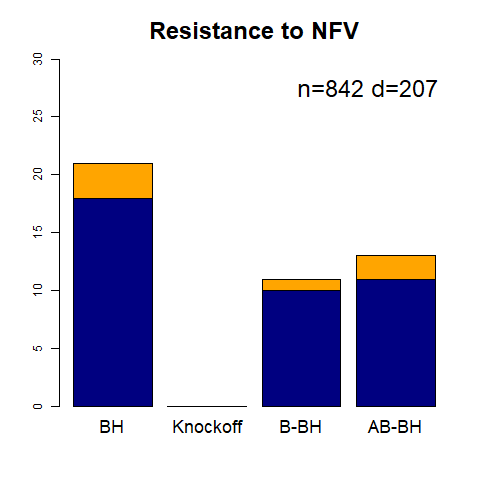}&
\includegraphics[width=0.28\textwidth]{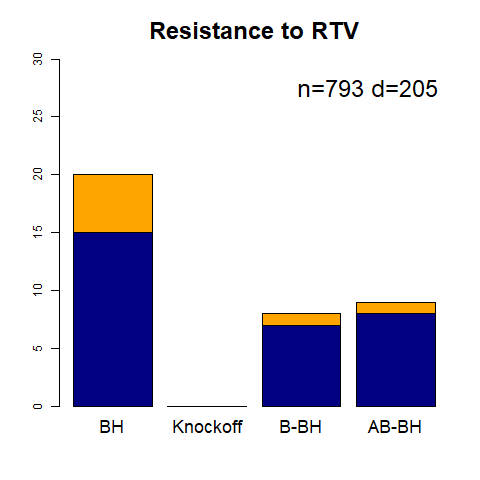}\\
\includegraphics[width=0.28\textwidth]{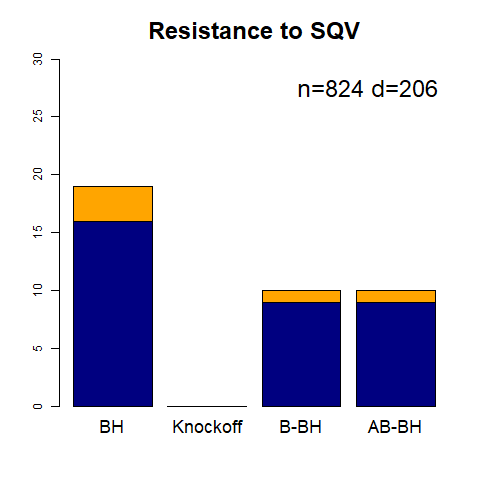}&
\end{tabular}

\caption{Real data example for $\alpha=0.05$.  %The horizontal line indicates the total number of HIV-1 protease positions appearing in the
%TSM list --
  %Note that the TSM list consists of mutations that are associated with the PI class of drugs %in general, and is not specialized to the individual drugs in the class. \\
Blue represents the number of discovers that are in the TSM list; and yellow represents the number of discoveries that not in the TSM list. The total number of HIV-1 protease positions in the TSM list is 34. \\
BH -- the BH method; Knockoff -- the knockoff based method of \cite{Barber2015}; B-BH -- the proposed Bonferroni-BH method; AB-BH -- the proposed adaptive Bonferroni-BH method.
\label{fig3} }
\end{figure}

We considered three different FDR levels, $\alpha=0.05$,  $\alpha=0.1$, and $\alpha=0.2$.  The results for $\alpha=0.05$ are reported in Figure \ref{fig3} and those for $\alpha=0.1$ and $\alpha=0.2$ are reported in Figures \ref{fig4} and \ref{fig5}, respectively, in Supplementary Material. We compare our proposed methods, the Bonferroni-BH and the Adaptive Bonferrni-BH, with the knockoff method of \cite{Barber2015} (that uses the data-dependent threshold $T$ in \eqref{eq:bct}) and the original BH method at level $\alpha$ using  $\hat{\beta}_{\text{OLS}}$, with the latter two being replicated from the study in \cite{Barber2015}.

The findings are summarized as follows. For the ATV drug, our methods generally  make no or few discoveries for all FDR levels. This might be due to the uncertainty level in the estimation of the parameters in these testing methods being relatively too high to detect more signals, since the sample size for this group is small, less than half of the sample sizes for most of the other six groups. No discovery is also the case for the knockoff method when $\alpha=0.05$ for the ATV drug. For the other six drugs, our proposed Bonferroni-BH and Adaptive Bonferroni-BH approaches identify reasonable number of discoveries, and the validations with the TSM reference also seem reasonable.
More specifically, we  observe that our methods are stable for  $\alpha=0.05, 0.1$ and $0.2$. That is, the Bonferroni-BH and the Adaptive Bonferroni-BH are making consistently more discoveries with increasing level of  FDR control, and the portions of the relative false discoveries with reference to the TSM list are also consistent.
On the other hand, the knockoff method is not seen to offer any discovery for six out of seven drugs at $\alpha=0.05$.  %and for one out of seven drugs at $\alpha=0.1$.
A plausible reason is that the HIV-1 data set is likely to contain relatively sparse signals (less than 15\% if referred to the TSM list), in which case, as our simulations above indicated, the Knockoff method does not perform well with lower FDR levels. Of course, when $\alpha=0.2$, the knockoff method makes more discoveries than our methods do.

%For  another level $\alpha=0.1$,  our method perform consistently.  Meanwhile, the knockoff method is making more discoveries, while there is still one case for RTV that the knockoff method does not reject any hypothesis at the FDR level $0.1$.

\section{Concluding Remarks}\label{s6}
This article contributes to the development of new line of research cross-fertilizing two seminal ideas on multiple inference used in modern statistical investigations -  use of $p$-value based multiple testing method to control false discoveries \citep{Benjamini1995} and use of  knockoff copy of the design matrix for variable selection in linear regression settings \citep{Barber2015}. Specifically, we introduce novel knockoff-assisted, $p$-value based FDR controlling methods for variable selection in the setting of fixed-design multiple linear regression when sample size is as large as the number of explanatory variables. Underpinning the novelty of our methods is a new understanding of the knockoff's role in setting the stage for the underlying multiple testing problem. The technical novelty comes from developing new multiple testing methods under this setting with proven control of the FDR under Gaussian noise fully capturing the correlation information on explanatory variables. Our  main idea of using knockoffs or generating additional variables to create new settings to successfully control the FDR can have applications  outside the scope of this paper. For instance, we can develop FDR controlling methods for multiple testing of Gaussian means with arbitrary but known covariance matrix differently from \cite {Fithian2020} and \cite{Fan2012}. Also, for this same testing problem, we can consider developing methods controlling other overall measures of type I errors, such as FWER (familywise error rate), or $k$-FWER (generalized FWER), or $k$-FDR (generalized FDR). See \cite{Lehmann2005} and \cite{Sarkar2008} for $k$-FWER and \cite{Janson2016}) for the use of $k$-FWER in the context of knockoff assisted variable selection, and \cite{Sarkar2007} for $k$-FDR.

Our proposed methods will continue to work when $X$ is random, irrespective of its distribution, if the conditional distribution of $Y$ given $X$ can be modeled as in (\ref{Model1}).

On adapting our methods to high-dimensional settings, there are several possibilities. One promising approach would be to develop a two-stage procedure following the line of work in \cite{Meinshausen2009} on controlling the FDR using $p$-values in high-dimensional variable selection based on data-splitting. More specifically, we can consider splitting the data into two parts of equal size. On one part, we can perform variable selection and dimension reduction using one of the existing methods. \cite{Meinshausen2009} gave a list of such methods including the Lasso, the adaptive Lasso, orthogonal matching pursuit, and sure independence screening, which ensure that all important explanatory variables are retained in the selection process under some regularity conditions. Some unimportant variables are allowed to be selected as long as the number of selected variables is less than $n/4$, so that knockoff copies can be created only for the selected variables using the other half of the data. Thus, our proposed knockoff-adjusted BH methods can be applied in the second stage for final selection of important variables. The symmetry of the two parts of the data allows us to implement this two-stage procedure twice, opening up the possibility of developing newer methods combining both results. We would like to pursue this line of research, along with investigating the data-splitting strategy in handling high-dimensional models, in our future work.

%As final remarks, we like to make the following points that highlight the take-home messages from our current research. The distribution-free approach in \cite{Barber2015} may not always perform satisfactorily, as indicated by our simulation study. Plausible explanations for this are conservative estimation of the FDP using the knockoffs and its high level of variability for stringent, but commonly used, FDR levels. In contrast, our approach to using the knockoffs judiciously under Gaussian noise assumption to create a paired-multiple testing framework that leads to BH-type FDR controlling methods is more advantageous in detecting sparse signals and handling stringent FDR levels. We strongly recommend using our approach in those situations.

\section{Acknowledgements} We are grateful to two anonymous referees and the Editor whose valuable suggestions and comments have greatly improved the presentation of our manuscript. We are also thankful to Edo Airoldi and William Fithian for their comments on an earlier version of this paper.

\end{document}